\documentclass[prb,aps,onecolumn,floatfix,manuscript]{revtex4}

\usepackage{dcolumn}
\usepackage{amssymb}
\usepackage{graphicx}
\usepackage{color}

\newcommand{\be}{\begin{equation}}
\newcommand{\bea}{\begin{eqnarray}}
\newcommand{\ee}{\end{equation}}
\newcommand{\eea}{\end{eqnarray}}

\begin{document}

\title{Random barrier double-well model for resistive switching in tunnel barriers}
\author{Eric Bertin$^1$, David Halley$^2$, Yves Henry$^2$, Nabil Najjari$^2$, Hicham Majjad$^2$, Martin Bowen$^2$,Victor DaCosta$^2$, Jacek Arabski$^2$ and Bernard Doudin$^2$ }
\affiliation{$^1$Universit\'e de Lyon, Laboratoire de Physique, ENS Lyon, CNRS, 46 All\'ee d'Italie, F-69007 Lyon\\
$^2$IPCMS, 23 rue du Loess, BP 43, F-67034 Strasbourg Cedex 2, France\\
}

\date{\today}

\begin{abstract}

The resistive switching phenomenon in MgO-based tunnel junctions is attributed to the effect of charged defects inside the barrier. The presence of electron traps in the MgO barrier, that can be filled and emptied, locally modifies the conductance of the barrier and leads to the resistive switching effects.
A double-well model for trapped electrons in MgO is introduced to theoretically
describe this phenomenon.
Including the statistical distribution of potential barrier heights for these traps leads to a power-law dependence of the resistance as a function of time, under a constant bias voltage. This model also predicts a power-law relation of the hysteresis as a function of the voltage sweep frequency.
Experimental transport results strongly support this model and  in particular confirm the expected power laws dependencies of resistance. They moreover indicate that the exponent of these power laws varies with temperature as theoretically predicted.

\end{abstract}

\maketitle

\section{Introduction}

Resistive switching effects \cite{revue} have been studied since the $70s$ in range of insulating oxides such as TiO$_{2}$ or Al$_{2}$O$_{3}$ for instance\cite{switchannees70, switchannees70bis,switchannees70NIO,switchannees70TIO}. This interest has been renewed for a few years, since giant and reproducible effects were observed in perovskites \cite{bednorz} as for example SrTiO$_3$ doped with chromium. It makes these materials good candidates for a new generation of memories. Yet, the underlying physical mechanisms are still unclear and different hypotheses have been put forward. Electro-migration of dopants or oxygen vacancies along filaments could reversibly create conducting path across the insulating layer \cite{filaments,szot}. Another hypothesis \cite{revue} suggests the accumulation of charges at  the electrode/insulator interface, which depends on the applied bias, and thus changing the Schottky barrier height.

The switching effect is in most cases studied on relatively thick films, on the order of 100 nm thick, but it has also been observed in some systems with a thin barrier allowing tunnel transport \cite{bowen,Freitas,FeCoB, APLnous,Doudin,MgO}. We showed for instance that MgO tunnel barriers with a few atomic layers of chromium \cite{APLnous} or vanadium \cite{PRBnous} at the MgO interface exhibited reproducible switching effects in Fe/Cr/MgO/Fe or Fe/V/MgO/Fe systems. This was attributed to the creation of oxygen vacancies in MgO at the interface with these "dusting" layers. These defects locally open  extra channels in parallel with the "standard" tunnel transport through the MgO barrier. Moreover, these systems exhibit an interesting behavior, with a relaxation of the conductance on long time scales --on the order of minutes. We  indeed observed logarithmic relaxation of the conductance under a constant bias voltage $U$. A strong influence of the voltage sweep frequency on the hysteresis in $I(U)$ curves was also observed.

	In a recent article \cite{PRBnous}  we proposed a phenomenological model accounting for  the relaxation of the conductance with time and the effect of sweep frequency on the hysteresis in $I(U)$ curves. This model did not discriminate between both hypotheses: either electro-migration of oxygen vacancies could create local conducting paths in the barrier, or the accumulation of charges inside the barrier could modify its potential height and thus the tunneling transport. Within the scope of this general model, both hypotheses could lead to the same mathematical expressions. In this model, the relaxation of conductance with time was expected to be exponential and not logarithmic.

	The present article partly justifies this phenomenological model and supports the hypothesis of electron trapping as the microscopic origin of the modification to the tunneling conductance. We indeed observe telegraphic noise in the conductance which is interpreted as the sign of electron trapping and untrapping due to its low activation energy. We then propose a double-well model to account for these trapping events.  In this new model, double wells are characterized by a random barrier height between the two trap states. Assuming an exponential tail for the distribution of these barrier heights, we obtain a power law dependence of the current hysteresis  $\Delta I $  as a function of the voltage sweep frequency of $I(U)$ curves.
Moreover, the exponent of the power law obtained in this model is proportional to the temperature. We performed systematic measurements as a function of temperature which clearly confirm these theoretical predictions.

\section{Evidence of electron trapping}

\subsection{Telegraphic noise}

\begin{figure}[!htb]
    \centering
        \includegraphics[width=0.4\textwidth]{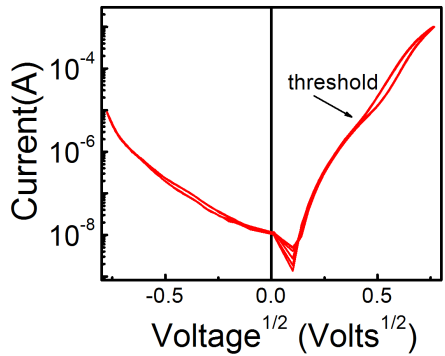}
        \caption{\label{Fig1}
       (color online) $I(U)$ curve measured at 80K showing hysteresis above +170 mV and a low current under a negative applied voltage.
                }
\end{figure}

    We  present results obtained on a Fe(20nm)/V(1.2nm)/MgO(3nm)/Fe(5nm)/Co(15nm) sample grown by molecular beam epitaxy. Details of the growth are given in Ref.~[15] together with the details of the micron-sized junction processing. Electrical measurements are performed with a conventional four-point DC technique. The reference of positive voltage was taken as the top electrode (with no vanadium). $I(U)$ curves on such samples were already shown at room temperature in Ref.~[15] and exhibited systematic hysteresis.
    At low temperature, the $I(U)$ curves are still hysteretic as can be seen in Fig. 1. We have to note a threshold (in the order of + 170 mV) below which the hysteresis is absent. Moreover, the junctions behave as a rectifier at low temperature: the current under negative bias becomes much smaller than under positive bias. This might be attributed to the asymmetric potential barrier in presence of the vanadium layer.     Indeed, the tunnel transport through monocrystalline MgO barrier is dominated by electrons having the  $\Delta1$ symmetry \cite{But2001} for which vanadium represents a large potential barrier \cite{Greu,Alv1978} - more than 4.2 eV-  due to its band structure. It thus leads to an asymmetric barrier with rectifying characteristics.   This point has to be further    studied. In the following, we will just show results obtained at a positive bias.   We observe for a low constant bias voltage (less than the threshold value of +170 mV), a telegraphic noise (see Fig. 2(a)), proving a bistable conductance of the junction. Provided that the voltage is lower than the threshold, almost no relaxation of the conductance value is observed on long time scales: the average level of conductance remains almost constant.
We have to stress the fact that Fe/MgO/Fe samples which do not show resistive switching effects or relaxation with time \cite{PRBnous}  do not show this telegraphic noise either. This supports the idea of a correlation between this observed noise and resistive switching mechanisms.

\subsection{Energy levels of electronic traps}

In the case of samples showing telegraphic noise, by slightly changing the applied dc bias U, we modify the occupancy rates between  the low and high resistance states labeled respectively (1) and (2).  This enables us to plot the ratio of occupancy states $\frac {\tau_2}{\tau_1}$ as a function of U and to fit this ratio as an exponential dependence on U (see the linearity on the Log plot of Fig.~2(b)). Supposing that the time occupancy for both states follows an Arrhenius law \cite{Ansermet,Son2010}  we obtain that
\be \label{DeltaE}
\frac{\tau_2}{\tau_1}=K e^\frac{\Delta E}{k_BT},
\ee
where $\Delta E= E_1-E_2$, K is a constant, $k_B$ is the Boltzmann constant
and $T$ is the temperature.
The energy of both states can moreover be written (see Fig.~4): $E_1=E_1^0+ \alpha U $ and $E_2=E_2^0- \alpha U $
where U is the applied voltage and $E_i^0$ is the energy of state $i$ in the absence of applied voltage. Thus, from the slope of $k_B T \ln(\frac{\tau_2}{\tau_1}) $ we can extract $\frac{d \Delta E}{d U}= 2 \alpha $ which corresponds to the voltage-dependent part of the energy difference between the two states. This yields $\frac{d \Delta E}{d U}=135$ meV/Volt for the measurement made at 80K. We found similar values on other junctions of the same sample.

By extrapolating our plot of Fig.~2 to $U=0$, we can have access to
$E^0\equiv E_2^0-E_1^0$ provided we make the hypothesis that $K=1$.
We then find that $E^0= -19$ meV.

 This value and its voltage dependant part $\alpha$ are very low compared to reported values obtained thanks to telegraphic noise in other  devices \cite{rall1991}: for instance $E_0 = 250 meV$ in Cu doped SeGe resisting switching systems \cite{Son2010}.  It is all the more striking as our values are obtained in MgO in which atomic displacements require high activation energies \cite{Vit1992,Alo2010}in the order of 2 eV. This is two orders of magnitude higher than what we observe. It thus supports the hypothesis of charge trapping instead of atomic displacement in order to explain the resistive switching in our system.

It is tempting to make the same type of measurements as a function of temperature. Unfortunately, obtaining such a dataset as a function of T was not possible: in some cases, no telegraphic noise is observed after increasing the temperature and stabilizing it -which takes more than 15 minutes- . Moreover, we can still observe such telegraphic noise for the new temperature, but associated to other metastable states.
This problem is not so crucial when changing the voltage at a given temperature: this is made rapidly, by small voltage steps, thus "following" the two states. Notice that the behavior shown on figure 2(b) is an indication that the same two states are considered when sweeping the voltage. This behavior was not observed as a function of temperature, suggesting that the telegraphic noise was associated to different states when modifying the temperature.

\begin{figure}[!htb]
    \centering
        \includegraphics[width=0.4\textwidth]{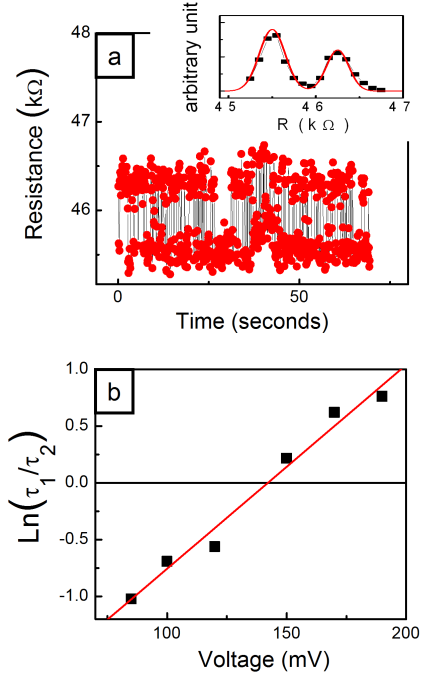}
        \caption{\label{Figbruit}
       (color online)(a) Resistance measurement at 80K under a constant voltage U=+120mV showing telegraphic noise.
       This is characteristic of a bi-stable single defect oscillating between two states with two different conductance values.       Inset: Histogram of the curve showing both populations on states 1 and 2 with two gaussian fits. (b) (squares) Experimental ratio of  residence times as a function of voltage, calculated from histograms. $\tau_2$ corresponds to the high resistance state. (full line) Linear fit.
                }
\end{figure}

This voltage dependence suggests that charges can be trapped in the MgO barrier or at the interface:  whether they are trapped or untrapped, the potential height of the tunnel barrier might be modified, thus influencing the  probability for electrons to tunnel from one electrode to another.
A similar phenomenon has already been observed in MOSFETs \cite{Ansermet} channels below the grid insulating oxide, or in thin Josephson \cite{Josephson} junctions. In both cases, oxygen vacancies in the oxide create charged defects whose charge fluctuates over time, leading to telegraphic noise in electronic transport as observed in our case.

In our case we can suppose that the electric charging of the barrier locally modifies its potential height due to electrostatic effects and thus changes the tunnelling probability of electrons close to this trap, as in the case of Coulomb Blockade. This effect might lead to strong conductance changes when the trap is located on a hotspot: it was indeed shown \cite{Dac2000,DacV2000,Her2010} that the tunnel transport through such thin barriers is not homogenous but dominated by some hotspots. They can for instance be due to a locally thinner barrier because of the roughness of the oxide. This rare events effect explains that a single electronic trapping could lead to 2 \%  changes of the junction conductance, provided it is located on such a spot.

Moreover, the value obtained for $\frac{d \Delta E}{d U}$ can be linked with the position of the defect inside the MgO barrier. Let us indeed suppose that the involved charged particle is an electron and that one trap position is at the V/MgO interface and the other at the MgO/Fe interface. The change of potential for the electron moving from one trap to another would be $q\cdot U$ where $q$ is the electron charge and $U$ the applied voltage. The value of $\frac{d \Delta E}{d U}$ should then be $1$eV/Volt which is higher than what we observe. This proves that the two traps are not located at the junction interfaces of the MgO barrier, but rather at a distance $d$ of each other with
\be
d= \frac{d \Delta E}{d U} d_{\rm MgO},
\ee
where $d_{\rm MgO}$ is the MgO barrier thickness. We find $d=0.4$nm. This would be consistent with the creation of oxygen vacancies at the lower interface close to the vanadium layer: $0.4$nm gives the order of magnitude of the thickness of the faulted MgO layer containing traps.

Looking at Fig.~2 we observe that a higher positive voltage favors the higher resistance state. Within our polarity convention, a positive voltage corresponds to electrons moving from the bottom --with vanadium-- interface to the top interface. In the case of trapping it thus suggests that a positive voltage favors charging the traps inside  MgO near the V/MgO interface,  and leads to a decrease of the conductance across the barrier.

Another point has to be stressed: at positive voltage the voltage-dependent part of the activation energy of traps and the constant part have opposite signs. Two regimes can thus be distinguished  relative to
\be
U_{th} = -E_0 \left(\frac {d E}{d U} \right)^{-1}
\ee
which is here of $140$mV. This value should be compared with the threshold value for hysteresis in $I(U)$ curves, close to $170$mV. Above this value, the voltage-dependent term dominates, leading to a partial filling of MgO traps  --and to an increase in the junction resistance. Below this voltage threshold value the constant term dominates, the trap states depopulate and the hysteresis disappears.

\section{Relaxation with time and role of the voltage sweep frequency}

We now turn to a higher voltage regime, i.e, with $U > + 170$mV. We showed in a previous article \cite{PRBnous} that the relaxation of the resistance under a constant voltage was nearly logarithmic at room temperature. Here we have performed  measurements of this relaxation at low temperature --from 10K to 200K. As seen  in the data of Fig. 3, the behavior remains the same at low temperature as shown on Fig.~3. We have to stress that no telegraphic noise is observed in this high voltage regime.  A Log-Log plot of the conductance G as a function of time yields a linear plot,
corresponding to a power-law with an exponent $m \ll 1$.
 This value strongly depends on the estimated value of $G_0$ which is the constant part of the conductance, corresponding to the asymptote on G(t) curves. For instance, as shown on Fig.~3 at 80K, $m\simeq 0.06 \pm 0.01$ if we take $G_0=1.2 .10^{-4} S$ , whereas we obtain $m\simeq 0.16 \pm 0.04$ if we take $G_0=3.6 .10^{-4} S$. Within this $[0.06-0.16]$ range all fits are correct.

  Nevertheless, this low value of $m$ indicates that the curves can be regarded to first approximation as almost logarithmic, as we did in Ref.~[15].

\begin{figure}[!htb]
    \centering
        \includegraphics[width=0.4\textwidth]{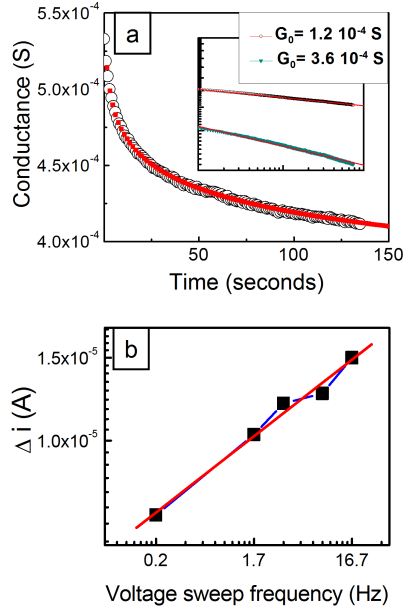}
        \caption{\label{FigLog}
      (color online) (a) (dots) Conductance G of the junction at 80K as a function of time under a +400 mV bias after applying for 3 minutes a -300mV voltage. (squares) Fit corresponding to $G=G_0 + a t^{-m} $ where $G_0=1.2.10^{-4} S$ is the evaluated non switching part of the conductance, $a$ is a constant and $m\simeq 0.06$. Inset: $G-G_0 = a t^{-m} $ on a Log-Log scale, with two different values for $G_0$ yielding two different exponent $m$ as explained in the text. (b) Hysteresis in current $\Delta I$ measured at +0.4V at 80K as a function of the bias sweep frequency. $\Delta I$ scales as $\omega^{m'}$ with ${m'}=0.22\pm 0.05$ (straight line).
                }
\end{figure}

Furthermore, in the case of dynamical measurements --i.e., when making $I(U)$ measurements-- we have shown \cite{PRBnous} that the hysteresis observed on $I(U)$ curves depends on the frequency of the voltage sweep: the $\Delta I(\omega)$ curve shows an increase at low frequency followed by a slight decrease at high frequency. If we focus on the low frequency regime (see Fig. 3b), $\Delta I $  follows a power law as a function of $\omega$; at $T=80$K, we find an exponent $m'\approx 0.22\pm 0.05$.

\section{Random barrier double-well model}

\subsection{Definition of the model}

\begin{figure}[!htb]
    \centering
        \includegraphics[width=0.4\textwidth]{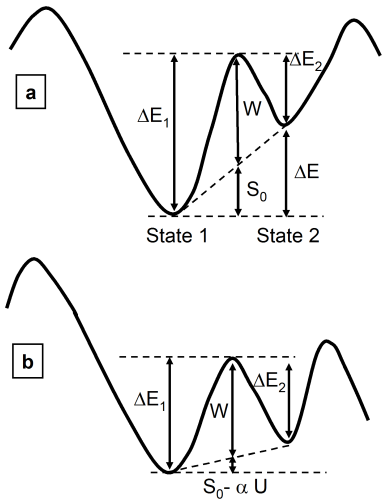}
        \caption{\label{Figwell}
       (a) Electronic potential landscape for an electron on two traps, without applied voltage. (b) With an applied positive voltage $U$.
                }
\end{figure}

The very slow relaxation of the conductance indicates that the system does not
possess a single relaxation time scale, but rather a broad distribution
of time scales, suggesting that disorder effects may play
an important role. It has been known for a long time that the presence
of disorder may strongly affect the electrical properties of materials,
for instance in ionic conductors like ``Hollandite'' \cite{Ber79},
or in amorphous insulating materials like As$_2$Se$_3$ \cite{Sch75,Pfi78},
and models based on random distributions of barrier heights
have proved useful to account for the behavior of such systems \cite{BG90,lee2009,vent2010}.
In this section, we propose a simple double-well model with a random
barrier between the two wells in order to describe our experimental data.

We develop this model under the hypothesis of electron charging of traps in MgO as suggested by experimental observations shown above. It would nevertheless give exactly the same mathematical results under the hypothesis of atomic drift of atoms, locally modifying the conductance. The double-well can indeed correspond to two positions of the involved ion inside the tunnel barrier, yielding two different values of the local tunnel conductance.

We consider that the electronic conductance $G$ results from many independent parallel conductance channels. Some of these channels are 'standard', and give altogether a contribution $G_0$ to $G$.
The other channels are modeled by double-well potentials in which electrons can be trapped.
These wells are assumed to be separated by an energy barrier of random height. In a given channel, the two potential wells have an energy $E_1={E_1}^0+\alpha U$ and $E_2={E_2}^0-\alpha U$ respectively ($E_1^0<E_2^0$), where $U$ denotes the electric potential and $\alpha$ is an effective electric charge.
To each well is also associated a given tunneling conductance, denoted as $g_1$ and $g_2$, with $g_2>g_1$.
For simplicity, we assume that $E_1^0$, $E_2^0$, $\alpha$, $U$, $g_1$ and $g_2$ have the same values in all the channels. In contrast, the energy barrier varies from one channel to another.
With an appropriate choice of the energy reference, we set $E_1^0=-S_0$ and
$E_2^0=S_0$. An electron going from the first well to the second one has to cross an energy barrier
\be \label{DeltaE1}
\Delta E_1 = W+S_0-\alpha U,
\ee
which defines $W$ (see Fig.~\ref{Figwell}).
In the opposite direction, the energy barrier is
\be \label{DeltaE2}
\Delta E_2 = W-S_0+\alpha U.
\ee
Hence $W$ can be interpreted as the average barrier between the two wells. We consider $W$ as a random variable, and denote as $\rho(W)$ its probability distribution.

Assuming that a number $n_c$ of non-standard channels are present in
the system, the total conductance at time $t$ is given by
\be \label{eq-Gt0}
G(t) = G_0 + n_c \overline{p_1}(t)\, g_1 + n_c (1-\overline{p_1}(t))\, g_2,
\ee
where $\overline{p_1}(t)$ denotes the average occupancy rate of the well
of conductance $g_1$ (the average being performed on the different channels, or equivalently, on the statistics of the barrier $W$).
At a temperature $T$, the mean time to cross the barrier $\Delta E_j$ is given
by an Arrhenius law
\be
\tau_j=\tau_0\, e^{\Delta E_j/k_B T},
\ee
where $\tau_0$ is a microscopic time characterizing the vibrations at the bottom of the wells.
We introduce the occupancy rate $p_1(t,W)$ of the first well, given the barrier $W$. The average occupancy rate is then obtained by averaging over the
barrier $W$, namely
\be
\overline{p_1}(t) = \langle p_1(t,W) \rangle_W.
\ee
The evolution equation for $p_1(t,W)$ reads
\be
\frac{\partial p_1}{\partial t}(t,W) = -\frac{1}{\tau_1} p_1(t,W)
+ \frac{1}{\tau_2} (1- p_1(t,W)),
\ee
which can be rewritten, using Eqs.~(\ref{DeltaE1}) and (\ref{DeltaE2}) as
\bea \label{master-eq}
\frac{\partial p_1}{\partial t}(t,W) &=& \frac{1}{\tau_0}\, e^{-W/k_B T}\\
\nonumber
&\times& \left[ e^{S(t)/k_B T} - 2p_1(t,W) \cosh (S(t)/k_B T) \right],
\eea
with $S(t)=S_0-\alpha U(t)$.

\subsection{Response to an electric potential step}

The relaxation of $p_1(t,W)$ after a step in the electric potential
$U(t)=U_0 \Theta(t)$ (where $\Theta(t)$ is the Heaviside function)
is readily calculated, yielding
\be
p_1(t,W) = p_1^{st} + A\, \exp\left(-\gamma t\, e^{-W/k_B T}\right),
\ee
with $p_1^{st}$, $A$ and $\gamma$ given by
\bea
p_1^{st} &=& \frac{1}{1+e^{-2(S_0-\alpha U_0)/k_B T}}\\
A &=& \frac{1}{1+e^{-2 S_0/k_B T}} - \frac{1}{1+e^{-2(S_0-\alpha U_0)/k_B T}}\\
\gamma &=& \frac{2}{\tau_0} \cosh \left(\frac{S_0-\alpha U_0}{k_B T}\right).
\eea
Averaging over the barrier $W$ yields
\be
\overline{p_1}(t) = p_1^{st} + A \langle \exp(-\gamma t\, e^{-W/k_B T})\rangle_W.
\ee
The average of the exponential term reads
\bea
\langle \exp(-\gamma t\, e^{-W/k_B T})\rangle_W &=&
\int_{W_\mathrm{min}}^{\infty} dW \rho(W)\\ \nonumber
&& \qquad \times \exp(-\gamma t\, e^{-W/k_B T})
\eea
where $W_\mathrm{min}$ is the minimum value of the barrier $W$.
In order to compute explicitly this last average, we need to choose a
specific form for $\rho(W)$.
Following the standard literature on trap and barrier
models \cite{Monthus96,BG90}, we consider a distribution $\rho(W)$
with an exponential tail,
\be \label{tail-rho-V}
\rho(W) \sim C\, e^{-W/W_0}, \quad W \to \infty,
\ee
where $C>0$ is a constant.
Such a form can be justified for instance on the basis of extreme
value statistics \cite{BM97}.
If $\rho(W)$ is purely exponential, $C$ is given by
\be
C = W_0^{-1}\, e^{W_\mathrm{min}/W_0}.
\ee
Making the change of variable $z=\gamma t\, e^{-W/k_B T}$,
we obtain for large time $t$
\be
\langle \exp(-\gamma t e^{-W/k_B T})\rangle_W \approx
\frac{C\,\Gamma(\mu)\, k_B T}{(\gamma t)^{\mu}}
\ee
with $\mu=k_B T/W_0$, and where $\Gamma(x)=\int_0^{\infty} dy\, y^{x-1} e^{-y}$
is the Euler Gamma function.
Accordingly, we have
\be
\overline{p_1}(t) = p_1^{st} + \frac{A\,C\,\Gamma(\mu)\, k_B T}{(\gamma t)^{\mu}}.
\ee
From Eq.~(\ref{eq-Gt0}), we thus find that the conductivity $G(t)$ relaxes as
a power law $t^{-\mu}$ to its asymptotic value, with an exponent $\mu$
proportional to the temperature. If the temperature is small, namely
$\mu \ll 1$, then the relaxation is approximately logarithmic
over a significant time window.
We underline that, within our model, the evolution of resistance with time is driven by the populations of states 1 and 2 which varies continuously with time. The fluctuation of resistance with time being averaged on a large number of defects, no telegraphic noise is expected.

\subsection{Response to a periodic excitation}

We now turn to the case of a small periodic excitation
$U(t) = u_0 \cos(\omega t)$, such that $\alpha u_0 \ll k_B T$.
We first consider a single channel, with a fixed barrier $W$.
Starting from Eq.~(\ref{master-eq}), we look for a solution of the form
\be
p_1(t,W) = p_1^0 + \frac{\alpha u_0}{k_B T}\, p_1^1(t,W)
\ee
and we linearize Eq.~(\ref{master-eq}) with respect to the small
parameter $\alpha u_0/k_B T$. The zeroth order
equation yields
\be
p_1^0 = \frac{1}{1+e^{-2S_0/k_B T}}.
\ee
At first order in $\alpha u_0/k_B T$, we get
\bea \label{lin-eq}
\tau_0 e^{W/k_B T} \frac{\partial p_1^1}{\partial t} &=&
-2 p_1^1(t,W) \cosh\frac{S_0}{k_B T} \\ \nonumber
&-&
\left(\cosh\frac{S_0}{k_B T}\right)^{-1} \cos(\omega t).
\eea
We look for a sinusoidal solution of the form
\be
p_1^1(t,W)=\Re\left[B(W)e^{i(\omega t+\phi(W))}\right],
\ee
with a real $B(W)>0$.
Inserting this form in Eq.~(\ref{lin-eq}) yields for $\phi$ and $B$
\bea \label{tan-phi}
\tan \phi(W) &=& -\frac{\omega\tau_0\, e^{W/k_B T}}{2\cosh(S_0/k_B T)}
\qquad (\cos \phi <0)\\
B(W) &=& \frac{[\cosh(S_0/k_B T)]^{-1}}
{[4 \cosh^2(S_0/k_B T) + (\omega\tau_0)^2\, e^{2W/k_B T}]^{1/2}}.
\nonumber
\eea
We now wish to quantify the hysteresis observed in the plane $(I(t),U(t))$.
We choose a value $U_1$ of the electric potential, such that
$0<U_1<u_0$. In the time interval $-\pi/\omega<t<\pi/\omega$, there are
two times, $t_1<0$ and $t_2=-t_1$ such that $U(t_1)=U(t_2)=U_1$.
Then the current intensity difference $\Delta I \equiv I(t_2)-I(t_1)$
is a measure of the time variation of the conductance, since
$\Delta I = U_1 \Delta G$, with $\Delta G \equiv G(t_2)-G(t_1)$.
From Eq.~(\ref{eq-Gt0}), $\Delta G$ is given by
\be
\Delta G = n_c (g_2-g_1)[\overline{p_1}(t_1)-\overline{p_1}(t_2)].
\ee
To compute this last expression, we start by considering a single channel,
that is, a fixed value of $W$.
The difference $\Delta p_1(W) \equiv p_1(t_1,W)-p_1(t_2,W)$ can be easily
determined:
\be
\Delta p_1(W) = \frac{B\alpha u_0}{k_B T} [\cos(\omega t_1+\phi)-\cos(\omega t_2+\phi)].
\ee
Taking into account the relation $t_2=-t_1$, we get
\be
\Delta p_1(W) = \frac{2B(W)\alpha u_0}{k_B T} \sin\phi(W) \sin\omega t_2,
\ee
Evaluating $\sin\phi$ from
Eq.~(\ref{tan-phi}), we obtain
\be
\Delta p_1(W) = \frac{2B(W)\alpha u_0 \omega\tau_0\, e^{W/k_B T}(1-U_1^2/u_0^2)^{1/2}}{k_B T[4 \cosh^2(S_0/k_B T)+(\omega\tau_0)^2\, e^{2W/k_B T}]^{1/2}}.
\ee
To obtain the current difference for the whole sample, we need to average
over the energy barrier $W$:
\be
\Delta I = U_1 n_c (g_2-g_1) \langle \Delta p_1\rangle_W.
\ee
With the notations
\bea
b &=& 2\cosh(S_0/k_B T) \\
D &=& \frac{2 \alpha u_0}{k_B T \, \cosh(S_0/k_B T)}
\left(1-\frac{U_1^2}{u_0^2}\right)^{\frac{1}{2}}
\eea
we have
\be
\langle \Delta p_1 \rangle_W = D \int_{W_\mathrm{min}}^{\infty}
dW\rho(W)\, \frac{\omega\tau_0\, e^{W/k_B T}}{b^2+(\omega\tau_0)^2 e^{2W/k_B T}}.
\ee
Introducing the change of variable $x=\omega\tau_0\, e^{W/k_B T}$, we find
\be
\langle \Delta p_1\rangle_W = D \int_{\omega\tau_{\mathrm{min}}}^{\infty}
\frac{dx}{\beta x}\, \rho\left(k_B T\ln\frac{x}{\omega\tau_0}\right)
\, \frac{x}{b^2+x^2}
\ee
with $\tau_{\mathrm{min}} = \tau_0\, e^{W_{\mathrm{min}}/k_B T}$.
For $\omega\ll \tau_0^{-1}$, $k_B T\ln(x/\omega\tau_0)$ is typically large,
and one can use the
asymptotic expression (\ref{tail-rho-V}) of $\rho(W)$, yielding
\be
\langle \Delta p_1\rangle_W = \frac{DC}{\beta}\, (\omega\tau_0)^{\mu}
\int_{\omega\tau_{\mathrm{min}}}^{\infty}
\frac{dx}{x^{\mu} (b^2+x^2)}.
\ee
If $\mu<1$, the integral converges to a finite value when its lower bound
goes to zero, and we get that $\langle \Delta p_1\rangle_W$ scales
as $\omega^{\mu}$. The remaining integral can be computed exactly,
and we eventually obtain for the average current variation
\bea
\Delta I &\approx&
\frac{\pi \alpha C}{2^{1+\mu}\cos\frac{\pi\mu}{2}\,[\cosh(S_0/k_B T)]^{2+\mu}}
\\ \nonumber
&&\qquad \times
U_1 u_0 \left(1-\frac{U_1^2}{u_0^2}\right)^{\frac{1}{2}}
n_c (g_1-g_2)\, (\omega\tau_0)^{\mu}
\eea
so that $\Delta I$ also scales as $\omega^{\mu}$, in the regime $\omega\ll \tau_0^{-1}$ and $\mu<1$ of experimental interest.

\subsection{Comparison between model and measurements}

Both calculations, in the case of a constant applied voltage and in the case of a varying voltage, lead to a power-law dependence of the conductance
as a function of time in the first case and of frequency in the second.
The model is therefore in qualitative agreement with our observations shown on Fig.~3 of a power law dependency. The  experimental values obtained for the exponents $m$ and $m'$ should, according to the model, be equal, whereas they slightly differ -see Fig. 3-. Nevertheless, the large error bars on these experimental values make that they are still compatible with our model.

Moreover, the model supposes  that the exponent $\mu$ is equal to $k_B T/W_0$, ie proportional to the temperature T. Indeed, we find a linear relation between the exponent $m'$ and temperature (see Fig. 5).

\begin{figure}[!htb]
    \centering
        \includegraphics[width=0.4\textwidth]{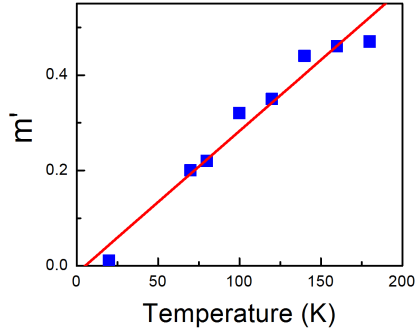}
        \caption{\label{Fig5}
       (color online) The power law exponent $m'$ as a function of T, obtained from fits of $\Delta I $ as a function of $\omega$. $\bigtriangleup I $ is evaluated at +400mV on a $\pm 0.6V$ $I(U)$ cycle. A linear fit of the data is given.
                }
\end{figure}

From our experimental observations we thus can roughly evaluate $W_0$, the "typical" value of the barrier height in the double-well model. Indeed, we have $\mu=k_B T/W_0$, so identifying $\mu$ to $m'$  gives ${W}_0= 57$meV.
This value is on the order of magnitude of the Coulomb Blockade energy for one electron in a tunnel barrier \cite{LivreShi}, which is yet another  argument in favor of a microscopic origin for the conductance modifications in terms of trapping and untrapping of electrons on defects in the barrier. This would be the origin of the resistance switching observed in our tunnel junctions.

We have to notice that the hypothesis made in the calculations, in the case of a periodic excitation, ie $\alpha u_0 \ll k_B T$, is not experimentally justified: $\alpha u_0 = 54 meV $ if we assess $\alpha$ at $0.135 eV/V$  and if we take $u_0= 0.4V$. This value is larger than $k_B T$ in the studied temperature range, which means that the linearized expression of Eq.~(\ref{lin-eq}) should be regarded as an approximation.

We cannot extract more quantitative information from the comparison with experimental observations:  our model does not predict the absolute value of the resistance relaxation with time, which would require for instance the knowledge of $g_1$,$g_2$ and $\tau_0$.

 To be complete, we note that,  on the one hand we observe telegraphic noise, thus associated to one defect, and on the other hand we model the junction in terms of a large distribution of defects, which could look contradictory.
In fact telegraphic noise is observed at low temperature and  low voltage, i.e., below $170$mV. For these values, we can suppose within our model that state 1 is dominant. It means that the traps inside the MgO barrier are empty, all but one. This leads to the telegraphic noise. It is sometimes observed with several levels thus involving different defects, yielding the addition of two telegraphic noise signals. On the contrary, at higher voltage or higher temperature, both populations  -i.e. electrons on states 1 and 2- are present and the telegraphic noise is smeared out due to contributions of many defects. This explains that we observe a continuous relaxation of conductance with time, without telegraphic noise. \\

\section{Conclusion}

In conclusion, we  showed here that a simple statistical model of electron trapping inside the MgO barrier could explain the resistance switching effects in MgO-based tunnel junctions. It also explains the long time relaxation of conductance according to a power-law behavior.
In addition, the temperature dependence of the theoretical exponent
is consistent with  experimental observations. Our model supposes a change of the tunneling probability of electrons due to a local charging of the barrier: in that sense, it differs from usual hopping models through traps.

We have to stress that this statistical model is in qualitative agreement with the phenomenological model that we proposed in Ref.~[15]: in this model, inspired from memristor models \cite{strukov},  we introduced an  electromigration term that makes the tunnel barrier height or thickness change as a function of the applied voltage, as well as an additional term which is voltage-independent. This second term makes the conductance relax towards a given value, independently of the applied voltage bias. Roughly speaking, this extra term plays a role analogous to that of thermal excitations in our present statistical model: thermal excitations indeed tend to equalize the populations of trapped and untrapped defects, and thus also tend to bring back the conductance to a given value, whatever the applied voltage.

Our present approach  is quite general for resistance switching effects in tunnel junctions: many resistance switching effects attributed for instance to ferroelectricity in the barrier --see for instance Ref.~[27]--  could perhaps be interpreted in terms of electron trapping on defects in the barrier.

M. Bowen acknowledges a financial support from the Agence Nationale de la Recherche (ANR-09-JCJC-0137).

\end{document}